\documentclass[
 reprint, prd, 
 amsmath,amssymb,
 aps,nofootinbib
]{revtex4-2}
\usepackage{graphicx}
\usepackage{dcolumn}
\usepackage{bm}
\usepackage{upgreek}
\usepackage{multirow}

\usepackage{graphicx}
\usepackage{textcomp}
\usepackage{xcolor}
\usepackage{physics}
\usepackage{caption}

\usepackage[ruled,vlined]{algorithm2e}

\SetCommentSty{mycommfont}

\usepackage{subcaption}
\usepackage{mathtools}

\definecolor{haykcolor}{HTML}{003dff}

\begin{document}

\preprint{APS/123-QED}

\title{A New Particle Pusher with Hadronic Interactions for Modeling Multimessenger Emission from Compact Objects}

\author{Minghao Zou}
 \email{minghao2006.zou@gmail.com}
 \affiliation{Valley Christian High School}

\author{Hayk Hakobyan}
\affiliation{Physics Department \& Columbia Astrophysics Laboratory, Columbia University, New York, NY 10027, USA}
\affiliation{Computational Sciences Department, Princeton Plasma Physics Laboratory (PPPL), Princeton, NJ 08540, USA}

\author{Rostom Mbarek}
\altaffiliation{Neil Gehrels Fellow}
\email{rmbarek@umd.edu}
\affiliation{Joint Space-Science Institute, University of Maryland, College Park, MD, USA}%
\affiliation{Department of Astronomy, University of Maryland, College Park, MD, USA}
\affiliation{Astrophysics Science Division, NASA Goddard Space Flight Center, Greenbelt, MD, USA}

\author{Bart Ripperda}
\affiliation{Canadian Institute for Theoretical Astrophysics, University of Toronto, Toronto, ON, Canada M5S 3H8}
\affiliation{David A. Dunlap Department of Astronomy, University of Toronto, 50 St. George Street, Toronto, ON M5S 3H4}
\affiliation{Department of Physics, University of Toronto, 60 St. George Street, Toronto, ON M5S 1A7}
\affiliation{Perimeter Institute for Theoretical Physics, Waterloo, Ontario N2L 2Y5, Canada}

\author{Fabio Bacchini}
\affiliation{Centre for mathematical Plasma Astrophysics, Department of Mathematics, KU Leuven, Celestijnenlaan 200B, B-3001 Leuven, Belgium}
\affiliation{Royal Belgian Institute for Space Aeronomy, Solar--Terrestrial Centre of Excellence, Ringlaan 3, 1180 Uccle, Belgium}

\author{Lorenzo Sironi}
\affiliation{Department of Astronomy \& Columbia Astrophysics Laboratory, Columbia University, New York, NY, 10027, USA}
\affiliation{Center for Computational Astrophysics, Flatiron Institute, 162 5th Avenue, New York, NY 10010, USA}
 
\date{\today}

\begin{abstract}
We propose novel numerical schemes based on the Boris method in curved spacetime, incorporating both hadronic and radiative interactions for the first time. Once the proton has lost significant energy due to radiative and hadronic losses, and its gyroradius has decreased below typical scales on which the electromagnetic field varies, we apply a guiding center approximation (GCA). We fundamentally simulate collision processes either with a Monte-Carlo method or, where applicable, as a continuous energy loss, contingent on the local optical depth. To test our algorithm for the first time combining the effects of electromagnetic, gravitational, and radiation fields including hadronic interactions, we simulate highly relativistic protons traveling through various electromagnetic fields and proton backgrounds. We provide unit tests in various spatially dependent electromagnetic and gravitational fields and background photon and proton distributions, comparing the trajectory against analytic results. We propose that our method can be used to analyze hadronic interactions in black hole accretion disks, jets, and coronae to study the neutrino abundance from active galactic nuclei.

\end{abstract}

\maketitle

\section{Introduction}
\label{sec:Introduction}

Recent IceCube neutrino observations show a significant flux in the direction of a single source, i.e., the Seyfert galaxy NGC~1068, with a large suppression of its expected $\gamma$ ray counterpart \citep{IceCube-NGC1068,padovani+24}. It has since been suggested that inner regions of
active galactic nuclei (AGNs) are astrophysical candidates to produce neutrinos, where plasma in combination with strong radiation fields contribute to the observed neutrino and electromagnetic flux \citep{murase+20,inoue+20,kheirandish+21,halzen+22,kurahashi+22,eichmann+22,fang+23,fiorillo+23b,mbarek+24}. 
As for the diffuse extragalactic component reaching PeV energies, other magnetically dominated neutrino sources have been proposed, such as compact regions within blazar jets \citep{murase+12, murase+14, rodrigues+21}. Although the significance level is low, a temporal association has been observed between the blazar TXS 0506+056 and a high-energy neutrino detection \citep{TXS065-18}. Beyond PeV energies, significant efforts are underway to detect neutrinos with energies $\gtrsim 10^{18}$~eV, which are expected to be linked to interactions of Ultra-High-Energy cosmic rays (UHECRs). Examples of such initiatives include GRAND \citep{GRAND20}, POEMMA \citep{POEMMA20}, IceCube-Gen2 \citep{ICECUBE-GEN2}, and PUEO \citep{PUEO21}.

Neutrinos can be emitted through interactions of protons with other particles or photons. For these interactions to be efficient, protons need to be accelerated to ultrarelativistic energies. Dilute and magnetized X-ray emitting coronae in the inner regions of AGNs are promising regions to accelerate protons and power the neutrino flux \cite{fiorillo+23b,mbarek+24,2024arXiv240701678F}.
In this context, a localized understanding of proton energization, radiative cooling, and scattering in strong electromagnetic and gravitational fields in the plasma surrounding supermassive black holes is essential. 



In this paper, we present a full charged particle propagation framework built upon a generalized Boris pusher, extending state-of-the-art numerical schemes for plasma-radiation interactions \cite{2019MNRAS.482L..60W,tristan_v2,2019ApJ...877...53H,2020ApJ...899...52S} and particle scattering \cite{Del_Gaudio_2020,2024PhRvL.132h5202G} focusing on protons in a curved spacetime \cite{Parfrey:2018dnc,2019ApJS..240...40B,2020ApJ...895..121C,Bacchini:2020}. The presented method includes a novel Monte Carlo scheme that captures discrete cooling effects from neutrino-producing hadronic processes, i.e., proton-proton ($pp$), photomeson ($p\gamma$), and Bethe-Heitler interactions (\emph{bth}). This generalizes the work presented in \citet{Mbarek:2022} where the source neutrino yield from UHECRs in AGN jets was investigated. To accurately handle small gyration scales compared to macroscopic scales of the black hole, we extend the method with a guiding center approximation (GCA) that is dynamically (de)activated in highly magnetized plasma regions \cite{Bacchini:2020}.

Our framework is designed to evolve particles in static or dynamically evolving electromagnetic and gravitational fields obtained from general relativistic magnetohydrodynamic (GRMHD), numerical relativity (NR), or general relativistic particle-in-cell (GRPIC) simulations\footnote{Here we assume the rate of temporal variations in gravitational fields is slower than that of spatial variations}. Our method enables us to study particle injection, propagation, confinement, and acceleration in turbulent plasma in black hole accretion disks, coronae, and jets; as well as how protons can power the observed neutrino and electromagnetic (e.g., X-ray or $\gamma$-ray) signals. The method generalizes to any spacetime, allowing applications for merging neutron stars or black holes, collapsing stars, as well as isolated pulsars and magnetars.

We provide test cases for the different aspects of the method, incorporating various gravitational and electromagnetic fields, scattering in photon and proton backgrounds, and radiative interactions, using both a generalized Boris pusher and GCA. The method requires prescribed (either analytic or obtained from a numerical solution) electromagnetic and gravitational fields, photon fields for $p\gamma$ and \emph{bth} interactions, and a proton background for $pp$ interactions. The separate components of the framework can be tested and generically implemented in existing (GR)MHD or (GR)PIC codes.


In section~\ref{sec:Solvers} we introduce the components of the particle evolution framework, starting with the Boris pusher for flat and curved spacetime, then the guiding center approximation coupling for particles with small gyroradii, the introduction of radiation drag terms, and we finish with a generic method for hadronic processes. In section~\ref{sec:tests} we introduce a number of known and new tests for the different components of the framework. We discuss the applicability of the method in different astrophysical settings in the conclusions in section~\ref{sec:astro}.

\section{Numerical Solvers}
\label{sec:Solvers}

Our main goal is to develop numerical algorithms to integrate the equations of motion for relativistic charged particles (in particular, protons, for studying neutrino emission) in the curved spacetime around compact objects. Since we aim to have algorithms capable of working with an arbitrary metric and coordinate system, further throughout the paper we will work in the 3+1 formalism \citep{Gourgoulhon_2007, Frauendiener3+1}, where the coordinate system is split into purely timelike and spacelike components, allowing us to employ the coordinate time $t$ as an independent variable and follow the evolution of positions and momenta with respect to $t$ independently of the choice of observer. In this notation, an arbitrary metric can be written in the form:

\begin{equation}
g_{\mu \nu} = \begin{pmatrix}
    -\alpha^2 + \beta_k \beta^k & \beta_i \\
    \beta_j & h_{ij}
\end{pmatrix},
\end{equation}
where $\alpha$ is the \emph{lapse function}, $\beta_i$ is the \emph{shift vector}, and $h_{ij}$ is the spatial part of the metric. Further in the paper we assume that the indices denoted with Latin letters ($i$, $j$, $k$, ...) vary from $1$ to $3$, while the Greek letters ($\mu$, $\nu$, $\lambda$, ...) go from $0$ to $3$ (i.e., include time).

In the 3+1 split formalism, the equation of motion for particles can be considered only for the spatial components of the four-velocity $u_\mu$. To account for all the complex physics which affect the motion of protons, we decompose the effective force acting on each particle into several components:
\begin{equation}
    \label{eq:momUpdate}
    m_p c\frac{d u_i}{dt} = \left(F_L + F_{\rm grav} + F_{\rm sync} + F_{\rm hadr}\right)_i,
\end{equation}
where $u_i=p_i/(m_p c)$ is the spatial part of the covariant four-velocity of the proton ($p_i$ being the spatial components of the covariant four-momentum, and $m_p$ the proton mass), $F_L$ -- the Lorentz force due to collective large-scale electric and magnetic fields, $D^i$ and $B^i$, $F_{\rm grav}$ -- the effective gravitational force, $F_{\rm sync}$ -- the drag forces due to synchrotron radiation, and $F_{\rm hadr}$ -- the drag force due to hadronic processes (namely considering $pp$, $p\gamma$, and $ bth$ interactions here). Time $t$ in this equation is the local time-coordinate measured by a fiducial observer (FIDO; for more details, see \cite{Gourgoulhon_2007}). Together with the position update, $dx^i / dt = \alpha u^i / \gamma - \beta^i$, where $\gamma = \sqrt{1+u_i u^i}$ is the Lorentz factor of the particle and $u^i = h^{ij}u_j$, the momentum update in~\eqref{eq:momUpdate} forms the full system for the equations of motion.

Below we detail how to numerically implement each of the four force terms separately, eventually providing a novel scheme to also incorporate hadronic interactions. 

\subsection{Lorentz force in curved spacetime}
\label{sec:LorentzForce}

We will employ the widely-used relativistic generalization of the Boris algorithm \citep{Boris.Shanny_1970,Ripperda:2017,ripperda2018a} to solve the Lorentz force numerically in a general coordinate system with orthonormal unit vectors. The Boris algorithm splits the full Lorentz force into two separate parts: two updates with half of the electric force, separated by an effective rotation due to the magnetic field. 

For plasma surrounding compact objects, typically the effective gyroradii of charged particles vary by orders of magnitudes both due to the variation of the local magnetic field strength, and due to particles being accelerated to ultra-relativistic energies. In those cases, it is computationally unfeasible to attempt to numerically resolve the gyroradius (and the gyrofrequency) with the Boris algorithm, and in many instances treating the motion of the guiding center of the orbit, while ignoring the gyration itself, is a more feasible and realistic approach. In these cases, we will integrate particle motion with a relativistic GCA method \citep{NORTHROP196179,VANDERVOORT1960401,Ripperda:2017}. The main advantage of a GCA approach is that it realistically captures highly magnetized particle motion (e.g., in jets or magnetospheres) alleviating the requirement to resolve the gyroradius and gyrofrequency. The GCA is however not valid when the particle Larmor radii are comparable with the curvature radius of the magnetic field, $|B/\nabla B|$, or in magnetic nulls, e.g., in magnetic reconnection layers.

A hybrid approach, coupling the full Boris scheme with the GCA depending on local conditions in the plasma, has been proposed by \cite{Bacchini:2020} and later successfully applied to pulsar magnetosphere simulations \cite{hakobyan2023}. Conditions for switching from one pusher to the other are typically based on the current gyroradius or gyrofrequency of the particle (e.g., comparing it to the spatial scale $\Delta x$ or the temporal scale $\Delta t$ that is resolved in the simulation), or the ratio of magnitudes of the electric $D$ and magnetic field $B$, switching to the Boris method in regions where the magnetic field vanishes or $D>B$. For our purposes, we further simplify the motion of the guiding center, by only maintaining the $\bm{D}\times \bm{B}$ drift term, i.e.\ generally the dominant drift in the GCA in highly magnetized regions.
To maintain general applicability in curved spacetimes, we employ an implicit method with a fixed-point iteration, which typically converges within a few iterations. The reason behind the fast convergence lies in the fact that the local magnetic field curvature radius is typically much larger than the characteristic displacement of a particle with speed $v$ within a single timestep: $|B/\nabla B| \gg v \Delta t$. Because of this, the computational expense of the algorithm is not significantly greater than an explicit solver (e.g., Runge-Kutta).


Details on the numeric discretization of both algorithms (Boris and GCA) in flat spacetime are presented in appendix~\ref{appendix:boris_gca}, while numerical tests are presented further in section~\ref{sec:tests}. Below we present the general routine for adapting these algorithms to curved spacetime following \cite{Parfrey:2018dnc,Bacchini:2020}.

\subsubsection{General relativistic Boris pusher}
\label{GR:Boris}

The gyration timescale of charged particles in strong magnetic fields is typically much smaller than the timescale associated with local spacetime curvature, so we can perform a Strang split to separate the particle motion update due to the electromagnetic field, the gravitational field, and the drag forces \cite{Parfrey:2018dnc,Bacchini:2020}. Thus, to generalize our schemes to curved spacetimes, we solve for the geodesic motion with an implicit method, while we employ the Boris method for the electromagnetic force. For the Lorentz push in both the Boris and the GCA regimes, we first transform to a locally flat spacetime and introduce a local orthonormal basis (different for each point in space), the \emph{tetrad} basis: $x^i\to x^{\hat{i}}$ \cite{Komissarov_2004}, which we denote with ``hatted'' Latin indices: $\hat{i}$. In this frame, the spatial part of the four-velocity, $u^{\hat{i}}$, and the electric and magnetic fields, $D^{\hat{i}}$ and $B^{\hat{i}}$, can be treated as in flat Cartesian spacetime.\footnote{Note, that in flat spacetime, where $\alpha = 1$, and $\beta^i =0$, $D^i$ reduces to $E^i$, where $E_i= \alpha h_{ij}D^j + \epsilon_{ijk}\beta^j B^k/\sqrt{h} $, where $\epsilon_{ijk}$ is the Levi-Civita symbol. For consistency, we will be using $D^i$ to indicate the electric field throughout the paper.} Further in the text, we use this fact to implement all the drag terms in the momentum update in~\eqref{eq:momUpdate} for the motion in generalized spacetime.

The motion of a charged particle can be described by \cite{Bacchini:2018zom,bacchini2019grlorentz,Bacchini:2020},
\begin{gather}
  \frac{1}{c}\dot{x}^i = \underbrace{\frac{\alpha}{\gamma} h^{ij}u_j -\beta ^i}_{V^i}  \label{eq:gr_xdot},\\
  m_p c\dot{u}_i = \left(F_{L} + F_{\mathrm{grav}}\right)_i \label{eq:gr_udot},\\
  \frac{F_{\mathrm{grav},i}}{m_p c} = -\gamma \partial_i \alpha + u_j \partial_i \beta^j - \alpha \frac{u_ju_k}{2\gamma} \partial_i h^{jk}.
\end{gather}
Here, $F_{L,i}$ is the covariant Lorentz force, and dots correspond to $d/dt$. Below, we will use the fact that the Lorentz force in the tetrad frame has exactly the same form as in flat spacetime:
\begin{equation}
    F_{L,\hat{i}} = q\alpha \left(D_{\hat{i}} + \epsilon_{\hat{i}\hat{j}\hat{k}} \frac{u^{\hat{j}}}{\gamma}B^{\hat{k}}\right)
\end{equation}

Ultimately, the algorithm for pushing a particle with the generalized Boris routine can be expressed as Algorithm \ref{alg:GR_boris}.

In steps 5 and 8, we implicitly resolve the push by iterating over $u_i^{**}$, and $x^{n+1,i}$ respectively, until the error converges. For the general spacetime, the relative error at iteration $k$ for both $u_i$ and $x^i$ can be defined as $\delta_k = \sqrt{|\Delta \hat{a}_i^k \Delta \hat{a}_j^k| h^{ij}}$, where $\Delta \hat{a}_i^k \equiv (a_i^k - a_i^{k-1}) / a_i^k$, and $a$ corresponds to either $u$ or $x$. In our experiments, the algorithm typically converges within $k\approx 3\mbox{--}4$ iterations up to $\delta_k\lesssim 10^{-8}$ (with double precision). For step 8, notice that the right-hand side of the $dx^i/dt$ equation is taken at timestep $n+1/2$, requiring the knowledge of the $u_i^{n+1/2}$, computed by transforming the $u_{\hat{i}}^{n+1/2}$. This latter transformation requires metric components at position $x^{n+1/2}=(x^{n+1}+x^n) / 2$, which is also updating per implicit iteration; thus, we couple the tetrad transformation itself into the implicit step. 

\begin{algorithm}
\caption{GR-Boris Routine}\label{alg:GR_boris}
\DontPrintSemicolon
\tcc{0. initial conditions $t=t^n$:}
$D^{n,i}$, $B^{n,i}$, $u_i^{n-1/2}$, $x^{n-1,i}$, $x^{n,i}$ \;
\tcc{1. interpolate fields to particle position:}
$D^i_p$, $B^i_p$ $\gets \lvert x^{n,i} \rvert \multimap$ $D^i$, $B^i$\;
\tcc{2. convert fields, velocity to tetrad basis:}
$D^{\hat{i}}_p$, $B^{\hat{i}}_p$ $\gets \lvert x^{n,i} \rvert \multimap$ $D^i_p$, $B^i_p$\; 
$u_{\hat{i}}$ $\gets \lvert (x^{n,i}+x^{n-1,i})/2\rvert \multimap $ $u_i$ \;
\tcc{3. flat spacetime Boris push with $\Delta t/2$:}
$u^{*}_{\hat{i}}$ $\gets \lvert D^{\hat{i}}_p$, $B^{\hat{i}}_p\rvert \multimap$ $u^{n-1/2}_{\hat{i}}$\;
\tcc{4. convert the velocity to covariant basis:}
$u^{*}_{i}$ $\gets \lvert (x^{n,i}+x^{n-1,i})/2\rvert \multimap$ $u^{*}_{\hat{i}}$\;
\tcc{5. update $u^{*}_{i}$ using eq. \eqref{eq:gr_udot} implicitly:}
$u_i^{**} = u_i^{*} + \frac{\Delta t}{m_p c} F_{\mathrm{grav}, i}(x^{n, i}, \left(u_i^{**}+u_i^{*}\right)/2)$\;
\tcc{6. convert the velocity to tetrad basis:}
$u^{**}_{\hat{i}}$ $\gets \lvert (x^{n,i}+x^{n-1,i})/2\rvert \multimap$ $u^{**}_{i}$\;
\tcc{7. flat spacetime Boris push with $\Delta t/2$:}
$u^{n+1/2}_{\hat{i}}$ $\gets \lvert D^{\hat{i}}_p$, $B^{\hat{i}}_p\rvert \multimap$ $u^{**}_{\hat{i}}$\;
\tcc{8. update the particle position implicitly:}
$u^{n+1/2}_{i}$ $\gets \lvert (x^{n+1,i}+x^{n,i})/2\rvert \multimap$ $u^{n+1/2}_{\hat{i}}$\;
$x^{n+1,i} = x^{n, i} + c\Delta t V^i(\left(x^{n+1,i} + x^{n, i}\right)/2,u^{n+1/2}_i)$\;
\end{algorithm}

\subsubsection{General relativistic GCA}\label{GR:GCA}

We can similarly implement the GCA in generalized spacetime by decomposing the velocity of the gyrocenter as a parallel component along the magnetic field and a $\bm{D}\times \bm{B}$-drift (while neglecting other drift terms). Since the $\bm{D}\times \bm{B}$-drift 3-velocity $v_{D,i} = \sqrt{h}\epsilon_{ijk}D^jB^k / B^l B_l$ is provided by the interpolated EM fields, it is sufficient to only update the parallel 4-velocity of the particle, $u_{\parallel} \equiv u_i \hat{B}^i$, where $\hat{B}^i \equiv B^i / \sqrt{B^lB_l}$. We can thus express the equations of motion as,
\begin{gather}
    \frac{1}{c}\dot{R}^i = \underbrace{\alpha \left(\frac{u_{\parallel}}{\gamma} \hat{B}^i + v_{D}^i\right) - \beta^i}_{\widetilde{V}^i}, \label{grgca_R}\\
    m_p c \dot{u}_{\parallel} = (F_L + F_{{\rm{grav}}})_i\hat{B}^i, \label{grgca_ups}
\end{gather}
where $R^i$ is the position of the gyrocenter. Note that here we only update the scalar parallel 4-velocity $u_{\parallel}$; however, we can reconstruct the full particle 4-velocity \cite{Bacchini:2020} as $u_i = u_{\parallel}\hat{B}_i + \gamma v_{D,i} + u_{\perp {g},i}$, where $u_{\perp {g},i}$ is the spatial part of the perpendicular 4-velocity of the gyration, $\gamma \equiv \kappa(1 + u_{\|}^2 + 2\mu B\kappa / (mc)^2)^{-1/2}$ is the Lorentz factor of the full velocity, and $\kappa \equiv (1 - v_{D, i}v_{D}^i)^{-1/2}$ is the Lorentz factor of the drift. The magnitude of $u_{\perp {g},i}$ can be recovered by conserving the first adiabatic invariant of the particle $\mu \equiv (m_p c)^2 u_{\perp g, i} u_{\perp g}^i / (2B\kappa)$ (the magnetic moment; $B\equiv \sqrt{B_iB^i}$), while its direction can be found by picking a random gyration phase (since gyration period is assumed to be much smaller than the timestep). 

We can split our routine into separate GR and electromagnetic pushes, with the full algorithm described in Algorithm \ref{alg:GC_GCA}.

\begin{algorithm}
\caption{GR-GCA Routine}\label{alg:GC_GCA}
\DontPrintSemicolon
\tcc{0. initial conditions $t=t^n$:}
$D^{n,i}$, $B^{n,i}$, $u_i^{n-1/2}$, $R^{n-1,i}$, $R^{n,i}$ \;
\tcc{1. interpolate fields to particle position:}
$D^i_p$, $B^i_p$ $\gets \lvert R^{n,i} \rvert \multimap$ $D^i$, $B^i$\;
\tcc{3. project $D^{i}_p$ and $u_{{i}}$ to $B^{i}_p$:}
$D^i_{\parallel} = D^{j}_p B_{p,j} B^i_p / (B_{p,l} B^l_p)$\;
$u_{\parallel,i} = u_{j} B_{p}^j B_{p,i} / (B_{p,l} B^l_p)$\;
\tcc{4. convert projections to tetrad basis:}
$D^{\hat{i}}_{\parallel}$ $\gets \lvert R^{n,i} \rvert \multimap$ $D^i_{\parallel}$\; 
$u_{\parallel,\hat{i}}$ $\gets \lvert (R^{n,i}+R^{n-1,i})/2\rvert \multimap$ $u_{\parallel,i}$ \;
\tcc{5. push the parallel velocity with $\Delta t/2$:}
$u_{\parallel,\hat{i}}^* = u_{\parallel,\hat{i}}^{n-1/2} + \frac{q\Delta t/2}{m_p c}D^{\hat{i}}_{\parallel}$ \;
\tcc{6. convert the velocity to coordinate basis:}
$u_{\parallel,i}^*$ $\gets \lvert (R^{n,i}+R^{n-1,i})/2\rvert \multimap$ $u_{\parallel,\hat{i}}^*$ \;
\tcc{7. compute the scalar parallel velocity:}
$u_{\parallel}^* = u_{\parallel,i}^*\hat{B}^i$ \;
\tcc{8. update the parallel velocity implicitly:}
$u_{\parallel}^{**} = u_{\parallel}^* + \frac{\Delta t}{m_p c}F_{\mathrm{grav}, i}(x^{n, i}, u'_i)\hat{B}^i$\;
$u'_i\equiv \frac{(u_{\parallel}^* + u_\parallel^{**})}{2}\hat{B}_i+ \gamma v_{D,i} + u_{\perp g,i}$\;
\tcc{9. convert projection to tetrad frame:}
$u_{\parallel,\hat{i}}^{**}$ $\gets \lvert (R^{n,i}+R^{n-1,i})/2\rvert \multimap$ $u_{\parallel,i}^{**}$\;
\tcc{10. push the parallel velocity with $\Delta t/2$:}
$u_{\parallel,\hat{i}}^{n+1/2} = u_{\parallel,\hat{i}}^{**} + \frac{q\Delta t/2}{m_p c}D^{\hat{i}}_{\parallel}$ \;
\tcc{11. update gyrocenter position implicitly:}
$R^{n+1,i} = R^{n,i} + c\Delta t \widetilde{V}^i((R^{n+1,i}+R^{n,i})/2, u_{\parallel,i}^{n+1/2})$\;
$u_{\parallel,i}^{n+1/2}$ $\gets \lvert (R^{n+1,i}+R^{n,i})/2\rvert \multimap$ $u_{\parallel,\hat{i}}^{n+1/2}$\;
\end{algorithm}

The error term in the implicit step 11 is computed similar to the previously described technique. In substep 8, since the equation is scalar, the error term at iteration $k$ can be explicitly computed as $\delta_k = |u_\parallel^k - u_\parallel^{k-1} / u_\parallel^k|$.

\subsection{Drag terms}

The dominant energy loss channels for protons in supermassive black hole magnetospheres are synchrotron radiation and hadronic interactions with other protons or background photons. 

\subsubsection{Synchrotron Radiation}



The effective synchrotron drag force in flat spacetime can be approximated using the classical Landau-Lifshitz formula~\cite{Landau:1975pou},
\begin{equation}
\begin{aligned}
\label{eq:syncdrag}
\bm{F}_{\text{sync}} &= \frac{2}{3}r_p^2 B_0^2 \left[\bm{\kappa}_R - \gamma\bm{u} \chi_R^2\right],~\text{where}\\
& \bm{\kappa}_R = \left(\bm{d} + \frac{\bm{u}}{\gamma} \times \bm{b}\right) \times \bm{b} + \left(\frac{\bm{u}}{\gamma} \cdot \bm{d}\right)\bm{d}, \\
& \chi_R^2 = \left|\bm{d} + \frac{\bm{u}}{\gamma} \times \bm{b}\right|^2 - \left(\frac{\bm{u}}{\gamma} \cdot \bm{d}\right)^2,
\end{aligned}
\end{equation}
where $r_p = e^2 / m_p c^2$ is the classical radius of the proton, $B_0$ is some arbitrary normalization of the electromagnetic field (to make the term in the square brackets dimensionless), while $\bm{d} \equiv \bm{D} / B_0$, and $\bm{b}\equiv \bm{B} / B_0$. Note that the fields, $\bm{D}$ and $\bm{B}$, as well as the four-velocity of the particle, $\bm{u}$, in this relation are taken in the locally flat orthonormal basis, i.e., $D^{\hat{i}}$, $B^{\hat{i}}$, and $u^{\hat{i}}$. Here we also implicitly neglect the terms $\propto \gamma (\partial/c\partial t + \bm{u}/\gamma\cdot \nabla) \{\bm{D},\bm{B}\}$, as we assume that the variations of electromagnetic fields over a single timestep are small (the validity of this assumption will become clear further).


For implementing this force in numerical simulations, it is further convenient to define a critical Lorentz factor, following \cite{UzdenskySpitkovsky_2014}, for which the fiducial acceleration force, $\eta_{\rm rec} B_0 |e|$, is exactly equal to the radiation drag force, $(2/3)r_p^2 B_0^2\gamma_{\rm syn}^2$ (for simplicity we approximate $\gamma |\bm{u}| \approx \gamma^2$, assuming the fiducial field is perpendicular; the coefficient $\eta_{\rm rec}\approx 0.1$ is a physically motivated choice\footnote{The choice of $\eta_{\rm rec}$ is based on the fact that in typical plasma-physical scenarios, the fastest acceleration is achieved in magnetic reconnection, resulting in an electric field strength of about $10\%$ of the reconnecting magnetic field strength.}). Following this definition of $\gamma_{\rm syn}$, we may further simplify the expression for the synchrotron drag force:

\begin{equation}
\label{eq:syncforce}
\frac{\bm{F}_{\text{sync}}}{m_p c} = \frac{\eta_{\rm rec} \omega_B^0}{\gamma_{\rm syn}^2}\left[\bm{\kappa}_R - \gamma\bm{u} \chi_R^2\right],
\end{equation}
where $\omega_B^0 = |e| B_0 / (m_p c)$ is the Larmor frequency for protons with $|\bm{u}|\approx 1$ moving perpendicular to the fiducial field of strength $B_0$. $\gamma_{\rm syn}$ provides an order-of-magnitude physical estimate on the critical Lorentz factor of protons, for which acceleration by an electric field $\sim \eta_{\rm rec} B_0$ may be hindered by synchrotron losses (the actual critical Lorentz factor, of course, depends on the ratio of the local magnetic field to $B_0$, and the pitch angle of the proton). Conveniently, the value of $\gamma_{\rm syn}$ only depends on the fiducial magnetic field strength, $B_0$:
\begin{equation}
    \gamma_{\rm syn} = \left(\frac{3}{2} \frac{e\eta_{\rm rec}}{r_p^2 B_0}\right)^{1/2} \simeq 5.5\cdot 10^{10}\left(\frac{B_0}{1~\text{G}}\right)^{-1/2},
\end{equation}
where the approximate expression assumes $\eta_{\rm rec} = 0.1$. This allows us to easily parametrize the cooling term in numerical simulations and rescale its strength to more tractable values, while still retaining the hierarchy of energy and timescales. With this notation, it also becomes clear why neglecting the temporal/spatial variation terms of the fields in the drag force is justified. Indeed, one can easily show that to the leading term in the dimensionless synchrotron drag force is $F_{\rm{sync}} / (\omega_B^0 m_p c) \sim \eta_{\rm rec} (\gamma / \gamma_{\rm syn})^2$, while the neglected term is of the order of $\sim \eta_{\rm rec} (\gamma/\gamma_{\rm syn})^2 (c/\Delta) / \gamma \omega_B^0$, where $\Delta \sim B_0/|(\partial/c\partial t + \bm{u}/\gamma\cdot\nabla)\{\bm{D},\bm{B}\}|$ is the characteristic temporal/spatial variation scale of local electric and magnetic fields. Thus, our assumption holds as long as $\Delta \gg c/\gamma\omega_B^0$.

Details on coupling this drag force to the classical Boris routine (both of which are performed in the locally flat orthonormal basis) can be found in \cite{Tamburini:2010}. Particles pushed in the GCA regime with small-enough Larmor radii are assumed to move strictly along the fieldlines with zero pitch-angle, and are not subject to synchrotron cooling.

\subsubsection{Hadronic Processes}
\label{sec:hadronic_drag}



Three interaction channels are important to consider to accurately evaluate the drag force of an energetic proton\footnote{These processes are also relevant for heavier nuclei, where a scaling with of interaction cross-sections with the atomic mass is necessary \citep{murase+20,mbarek+24}.}: $pp$, $p\gamma$, and \emph{bth}, all of which may slow down the proton significantly and produce secondary particles, including neutrinos ($\nu$), $\gamma$-rays, pairs ($e^{-}e^{+}$), neutrons ($n$), and muons ($\mu$). Specifically \cite{Mbarek:2022}:
\begin{equation}
\begin{aligned}
    pp:~&p + p \to p + n + \pi^{+}\\
    p\gamma:~&p + \gamma \to p + \pi^0 + (\pi^{+} + \pi^{-})\\
    bth:~&p + \gamma \to e^{-} + e^{+} + p \\
\end{aligned}
\end{equation}
The produced pions ($\pi$) further decay to produce potentially observable neutrinos ($\nu$) and $\gamma$-rays,
\begin{equation*}
    \begin{cases}
    \pi^{+} \to e^{+} + \nu_{\mu} + \bar{\nu}_{\mu} + \nu_e\\
    \pi^{-} \to e^{-} + \nu_{\mu} + \bar{\nu}_{\mu} + \bar{\nu}_e
    \\
    \pi^{0} \to \gamma + \gamma
    \end{cases}
\end{equation*}

In a numerical algorithm, either MHD, or PIC, these interactions can be modeled in three different ways, each with their own advantages and drawbacks:

\begin{itemize}
    \item as an \emph{effective continuous drag force}: where all the individual interactions are averaged into a single effective force term, applied at each timestep in the tetrad frame. This approach is computationally the cheapest; however, it does not give any information about the secondary products of interactions, and may furthermore fail when the mean-free-path of the interaction is comparable to the scale of the system in question;
    \item as a \emph{probabilistic drag}: where the average force at a given timestep is applied probabilistically; with this approach, one recovers the average behavior of the previous case with the same computational cost, while also allowing the study of the reaction byproducts and the propagation of protons without any interactions for long distances; the drawback is that it puts a constraint on the simulation timestep, as the interaction optical depth has to be $\lesssim 1$ at a distance of $c\Delta t$;
    \item as a \emph{Monte-Carlo two-body scattering}: where the proton probabilistically interacts with a random sampling of the background particles (either protons or photons); while being the most accurate, this approach is also the most expensive computationally. This algorithm is useful when modeling both the energetic particle, and the background from first-principles, with potential back-reaction of the interaction on the background; for the context of this paper, this approach is excessive, as in typical black-hole environments the most energetic protons constitute a small fraction of the total energy of the system.
\end{itemize}

Below we discuss our implementation of the hadronic interactions, simulating these as drag forces acting either continuously or probabilistically. We also present how interaction cross sections have been computed in all of the cases.

{\bf $pp$ interactions.} In the context of this interaction, we assume the evolved proton to be relativistic, with the background proton population having typical much smaller energies: $|\bm{u}_{p,b}|\ll |\bm{u}|$. The probability of a proton-proton interaction during a time interval $\Delta t$ can be written as $p_{pp} = \text{min} (n_{p,b} \sigma_{pp} c\Delta t, 1)$, where $\sigma_{pp}$ is the interaction cross section in the rest-frame of either the incoming or the background proton, and $n_{p,b}$ is the number density of background protons (we also define $\lambda_{pp}\equiv 1/(n_{p,b}\sigma_{pp})$). 
Defining fiducial value for the mean-free path, $\lambda_{pp}^0 \equiv 1 / (n_{p,b}^0\sigma_{pp}^0)$ (where $n_{p,b}^0$ and $\sigma_{pp}^0$ are the fiducially defined number density and cross section), we can further rewrite the probability in the dimensionless form:
\begin{equation}
    p_{pp} = \text{min} \left[\frac{c\Delta t}{\lambda_{pp}^0} \left(\frac{n_{p,b}}{n_{p,b}^0}\right)\left(\frac{\sigma_{pp}}{\sigma_{pp}^0}\right), 1\right],
\end{equation}
which allows us to rescale the value of the effective mean free path for implementing in simulations with reduced scale separation, while still retaining the hierarchy of scales. 

Notice that the probability written above is only true in the ``lab'' frame, where the background is at rest. To model interaction with a non-stationary background, we transform the velocity of the incoming proton, $\bm{u}\to \bm{u'}$, the number density of background protons, $n_{p,b}\to n_{p,b}'$, and the time step, $\Delta t \to \Delta t'$, to the rest frame of the background population. We then evaluate the cross section, $\sigma_{pp}'$, and the probability $p_{pp}'$ of interaction. We then either apply the drag term probabilistically or continuously per timestep, in the following way:
\begin{equation}\label{eq:pp_prob_cont}
    \begin{aligned}
        \textrm{probabilistic:}~& \textrm{if}~(\mathrm{rnd}<p_{pp}')\Rightarrow\bm{u}_{p,\textrm{new}}' = \bm{u}_p'(1-\xi_{pp}),\\
        \textrm{continuous:}~& \bm{u}_{p,\textrm{new}}' = \bm{u}_p'(1-p_{pp}'\xi_{pp}),\\
    \end{aligned}
\end{equation}
where rnd is a random number sampled uniformly between $0$ and $1$, and $\xi_{pp}\approx 0.17$ is the characteristic inelasticity of $pp$ scattering \citep{kelner+06}. We can then transform back to the ``lab'' frame, recovering the final velocity after the scattering $\bm{u}_{p, \mathrm{new}}$.

{\bf $p\gamma$ interactions.} 
The cooling time via $p\gamma$ interactions for a proton (with a Lorentz factor $\gamma_p$) in the frame where the background radiation (with a distribution function $dn_\epsilon/d\epsilon$) is isotropic (``lab'' frame) can be written as \cite{mbarek+24},
\begin{equation}\label{eq:pg_cool}
	t^{-1}_{p \gamma}(\gamma_p) = \frac{\xi_{p\gamma} c}{2 \gamma_p^2} \int\limits^{\infty}_{\bar{\epsilon}_{\rm th}/2\gamma_p} d\epsilon \frac{dn_{\epsilon}}{d\epsilon} \epsilon^{-2}  \int\limits^{2 \gamma_p \epsilon}_{\bar{\epsilon}_{\rm th}}  \bar{\epsilon} \bar{\sigma}_{p\gamma}(\bar{\epsilon}) d\bar{\epsilon},
\end{equation}
Here, $\xi_{p\gamma} \approx 0.2$ is the average inelasticity for $p\gamma$ interactions \citep{kelner+08}. $\bar{\epsilon}_{\rm th} \approx 0.145$ GeV is the threshold photon energy in the proton rest frame, $\bar{\epsilon}$ is the photon energy in the proton rest frame, and $\epsilon$ and $dn_{\epsilon}/d\epsilon$ are the energy and the distribution function of the photons in the ``lab'' frame. Here both $\epsilon$ and $\bar{\epsilon}$ are in units of $m_pc^2$. 

For a constant cross section $\bar{\sigma}_{p\gamma} = \sigma_{p\gamma}^0$ and a perfect black-body photon distribution with a temperature $T_\varepsilon$, one can obtain an analytical solution for the proton cooling time $t_{p\gamma}$,
\begin{equation}
    t^{-1}_{p \gamma} = \frac{2\xi_{p\gamma}\sigma_{p\gamma}^0 T_\varepsilon^3}{\hbar^3 c^2 \pi^2}(\mathrm{Li}_3(\exp{-\psi}) + \psi \mathrm{Li}_2(\exp{-\psi})), \label{eq:pg_cool_time_CMB}
\end{equation}
where $\psi = \bar{\epsilon}_{\rm th}/(2 \gamma_p T_\varepsilon)$, and polylogorithm function of order $s$ in $z$ $\mathrm{Li}_s(z) \equiv \sum_{k = 1}^\infty z^k/k^s$, where $k$ is a placeholder variable we sum over. For a more realistic cross section, we perform the integration numerically, and equation~\eqref{eq:pg_cool_time_CMB} serves to test our numerical integration with known analytical results. We present the comparison of the two cases in the next section. 

If the background radiation is not isotropic, we can always find a reference frame where the non-diagonal components of the stress-energy tensor vanish, and consider the interaction in that frame (see appendix~\ref{appendix:photon_field_trans}). Then, similar to the $pp$ interaction, we can implement $p\gamma$ drag both as a continuous and a probabilistic force, by defining the interaction probability at each timestep in that frame as $p_{p\gamma}'\xi_{p\gamma} = \Delta t' / t_{p\gamma}'$, where we can use \eqref{eq:pg_cool} to compute $t_{p\gamma}'$ in the background radiation frame. Then the algorithm can be written as:
\begin{equation}
    \label{eq:pg_force_alg}
    \begin{aligned}
        \textrm{probabilistic:}~& \textrm{if}~(\mathrm{rnd}<p_{p\gamma}')\Rightarrow\bm{u}_{p,\textrm{new}}' = \bm{u}_p'(1-\xi_{p\gamma}),\\
        \textrm{continuous:}~& \bm{u}_{p,\textrm{new}}' = \bm{u}_p'(1-\Delta t' / t_{p\gamma}'),\\
    \end{aligned}
\end{equation}
with a subsequent transformation of the updated proton velocity back to the ``lab'' frame.

{\bf $bth$ interactions.} 
In Bethe-Heitler, photons interact with an atomic nucleus (a proton, in this case) producing an electron-positron pair, and slowing the proton down. In this sense, the interaction is similar to photomeson, so we can calculate the cooling time (in the frame where radiation is isotropic) similarly as, 
\begin{gather}\label{eq:bth_cool}
    t_{bth}^{-1} = \frac{c}{2 \gamma_p^2} \int\limits^{\infty}_{\bar{\epsilon}_{bth} / 2 \gamma_p} d\epsilon \frac{dn_\epsilon}{d\epsilon} \epsilon^{-2} \int\limits^{2 \gamma_p \epsilon}_{\bar{\epsilon}_{bth}} \bar{\epsilon} {\xi}_{bth}(\bar{\epsilon}) \bar{\sigma}_{bth}(\bar{\epsilon}) d\bar{\epsilon},
\end{gather}
where, again, the barred quantities are evaluated in the proton rest-frame, while the rest is in the ``lab'' frame, and $\bar{\epsilon}_{bth}=2m_e c^2\approx 1$ MeV. Note that in this case, the dependence of the inelasticity on energy is significant (see appendix of \cite{1992ApJ...400..181C}), so the analytical integral does not treat it as a constant (like in $pp$ and $p\gamma$ cases). However, for simplicity in numeric integration, we can obtain an effective value for ${\xi}_{bth}\bar{\sigma}_{bth}$ by taking its maximum, $2.5r_e^2 \alpha_F Z^2m_e/(Am_p)$, where $r_e = e^2/(m_ec^2)$ is the classical electron radius, $\alpha_F \approx 1/137$ is the fine-structure constant, $Z$ is the atomic number, and $A$ is the atomic mass. This yields an effective value for protons ${\xi}_{bth}\bar{\sigma}_{bth} \approx 7.5 \cdot 10^{-31} \rm{cm}^{2}$ (consistent with \cite{murase+10}). From there, we can then obtain an analytical solution for the cooling time similar to equation \eqref{eq:pg_cool_time_CMB}. 

The continuous drag force acting on the proton can then be implemented similarly to $p\gamma$ interactions \eqref{eq:pg_force_alg}, with the value of $t_{bth}$ computed in the background radiation frame. To apply the force probabilistically, however, one has to compute the value of ${\xi}_{bth}$ at each timestep to recover the interaction probability, $p_{\rm bth} = (\Delta t / t_{bth}) / \xi_{bth}$. In this case, we can obtain approximate expressions of ${\xi}_{bth}$ for different ranges of $\gamma_p$ (see equations (3.6)–(3.10) of \cite{1992ApJ...400..181C}). However, in reference to the system length scales ($\propto r_g \equiv GM/c^2$ for gravitational constant $G$ and system mass $M$), the probabilistic scheme offers negligible corrections to the particle trajectory, so it is more efficient to directly implement the continuous force. 


{\bf Computing the cross sections.}
The cross sections of $pp$ and $p\gamma$ interactions depend on the target particle energy in the proton rest frame. We extracted raw experimental data from the Particle Data Group \cite{ParticleDataGroup:2020ssz} for both interactions and proceeded to fit the functions $\log_{10}\sigma_{pp}(\log_{10}\bar{\gamma}_p)$ and $\log_{10}\sigma_{p\gamma}(\log_{10}\bar{\epsilon})$, where $\bar{\gamma}_p$ is the target proton energy in the traveling proton rest frame, with a $k$-th order rational function $s_{pp,p\gamma} P_k/Q_k$ for $k$-th degree polynomials $P_k$ and $Q_k$ and scaling constant $s_{pp,p\gamma}$. Here, $k=4$ for $pp$ interactions and $k=6$ for $p\gamma$ interactions. Specific coefficient values of fitted functions and plots are detailed in appendix~\ref{appendix:cross_sections}. 

For \emph{bth} interactions, while an analytical expression for $\xi_{bth}\sigma_{bth}$ exists, it is ineffective to implement in a full simulation due to its complex nature (see equation (3.4) from \cite{1992ApJ...400..181C}). Thus, we employ a polynomial fit for $\xi_{bth}\sigma_{bth}$ as a function of $\log_{10}(\bar{\epsilon})$ to approximate its analytical expression. We use a 12th degree polynomial to fit $10^3$ analytical points from the range $10^{-3} < \epsilon' < 10^2$ since any value for $\xi_{bth}\sigma_{bth}$ outside this range is effectively zero. Again, the coefficient values are detailed in appendix~\ref{appendix:cross_sections}.

\section{Numerical tests}
\label{sec:tests}

In this section, we describe a set of numerical unit tests which verify the validity of the algorithms described in the previous section~\ref{sec:Solvers}. In these experiments, we compare the trajectories of particles with analytic predictions, and verify the convergence of the numerical scheme. For the tests modeling the hadronic drag, we compare the previously described continuous-force approach with the probabilistic one, and also ensure that the basic kinematics in an extreme case of high optical depth is recovered. 

\subsection{Motion in EM fields in flat spacetime}

Unit tests described in this section model the trajectories of protons in analytically defined electromagnetic fields. We include tests both for the motion with and without a synchrotron drag, all performed in flat Cartesian spacetime.

\subsubsection{$\bm{D} \times \bm{B}$-drift}

The $\bm{D} \times \bm{B}$-drift describes particle motion under perpendicular magnetic and electric fields, where the particle drifts in the $\bm{D} \times \bm{B}$ direction. In this test, we initialize a positively charged particle of mass $m$ and charge $q$ at $\bm{x} = \bm{0}$ with $\bm{u} = \bm{0}$. Following \cite{Ripperda:2017}, we set $\bm{B} = \hat{\bm{x}}B$ and $\bm{D} = \hat{\bm{z}}D$, where we choose $D<B$ such that the Lorentz factor of the drift, $\kappa = 1/\sqrt{1-v_D^2} \in \{10, 100\}$. We then vary $\Delta t$ to test the convergence of the scheme (in this case, the flat spacetime Boris pusher described in the appendix~\ref{appendix:boris_gca}). 

\begin{figure*}[htb!]
    \centering
    \includegraphics[width=\textwidth]{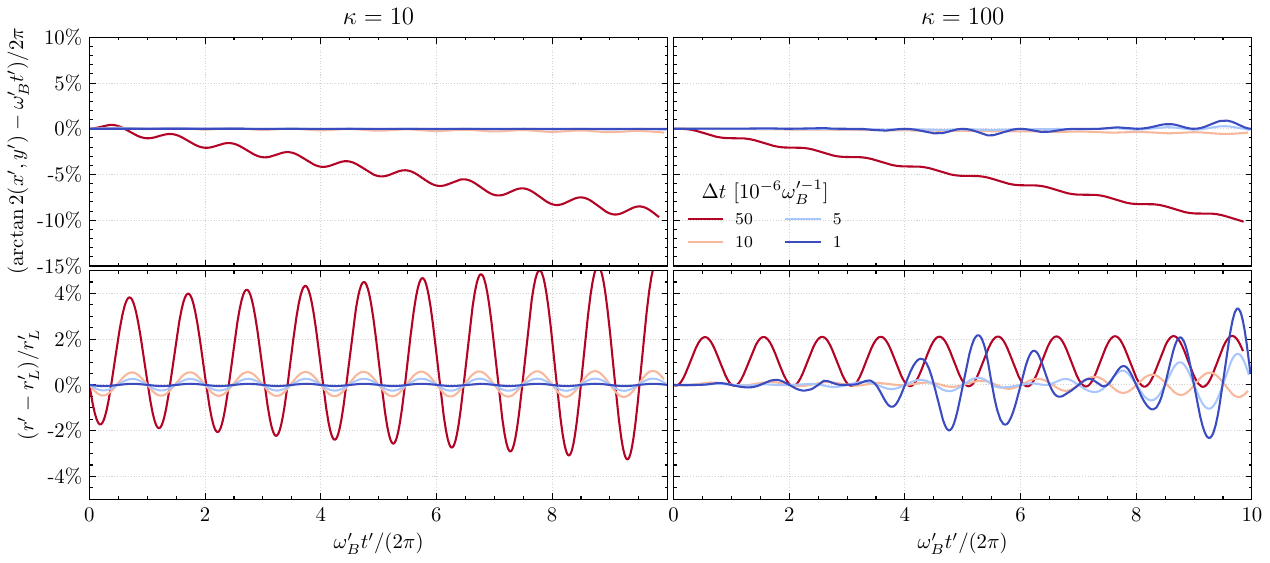}
    \caption{$\bm{D}\times\bm{B}$ drift simulation errors compared to analytical solutions for the Boris pusher in flat spacetime.}
    \label{fig:ExB}
\end{figure*}

The results are shown in figure~\ref{fig:ExB}, where the left two panels correspond to the case with $\kappa = 10$, and the right ones to $\kappa = 100$. To simplify the comparison with analytics, we plot all the quantities transformed to the $\bm{D}\times\bm{B}$ frame\footnote{The simulation itself is performed in the lab frame.}, where the motion is purely circular with a constant angular frequency, $\omega_B'\equiv \omega_B^0 / \kappa^2$, with $\omega_B^0 = qB/(mc)$. The top two plots show the relative error in the gyration phase with respect to $\omega_B' t'$, where $t'$ is the time in the $\bm{D}\times\bm{B}$ frame. The two plots at the bottom show the relative deviation of the particle gyroradius, $r'$, from the analytic prediction, $r_L'\equiv \kappa^2 v_D r_L^0$, with $r_L^0 = mc^2/(qB)$. Different colored lines correspond to different values of the timestep, which we vary in the range $\Delta t \in [10^{-6}, 5\cdot 10^{-5}]~\omega_B'^{-1}$.

The relative error in the gyroradius (measured in the drift frame) fluctuations lies within $1\mbox{--}2\%$ for $\Delta t\lesssim 10^{-5}\omega_B'^{-1}$, with the amplitude of its fluctuations converging with smaller timestep. Similarly, the phase lag error fluctuations are bound to within a $1\mbox{--}2\%$ of $2\pi$. The least time-resolved case, corresponding to $\Delta t = 5\cdot 10^{-5}\omega_B'^{-1}$, shows a slowly growing phase lag, while the error in gyroradius is still bound to within a few percent for both $\kappa=10$ and $100$. 

\subsubsection{Synchrotron cooling and transition to GCA}

In the following experiment, we simulate the trajectory of a proton which undergoes synchrotron cooling, while moving in a uniform magnetic field. We start with the same magnetic field, $\bm{B} = \hat{\bm{x}}B$, but in this case, the particle is initialized with its four-velocity $\bm{u} = \hat{\bm{x}}\tan{(\pi/3)} + \hat{\bm{y}}$ having a nonzero pitch angle with respect to the magnetic field. For the purposes of this test, we set the value $\eta_{\rm rec}/\gamma_{\rm syn}^2 = 1$, while the motion is described by a combination of the Lorentz force and the synchrotron drag force from equation~\eqref{eq:syncforce}.

\begin{figure}[htb!]
    \centering
    \includegraphics[width=\columnwidth]{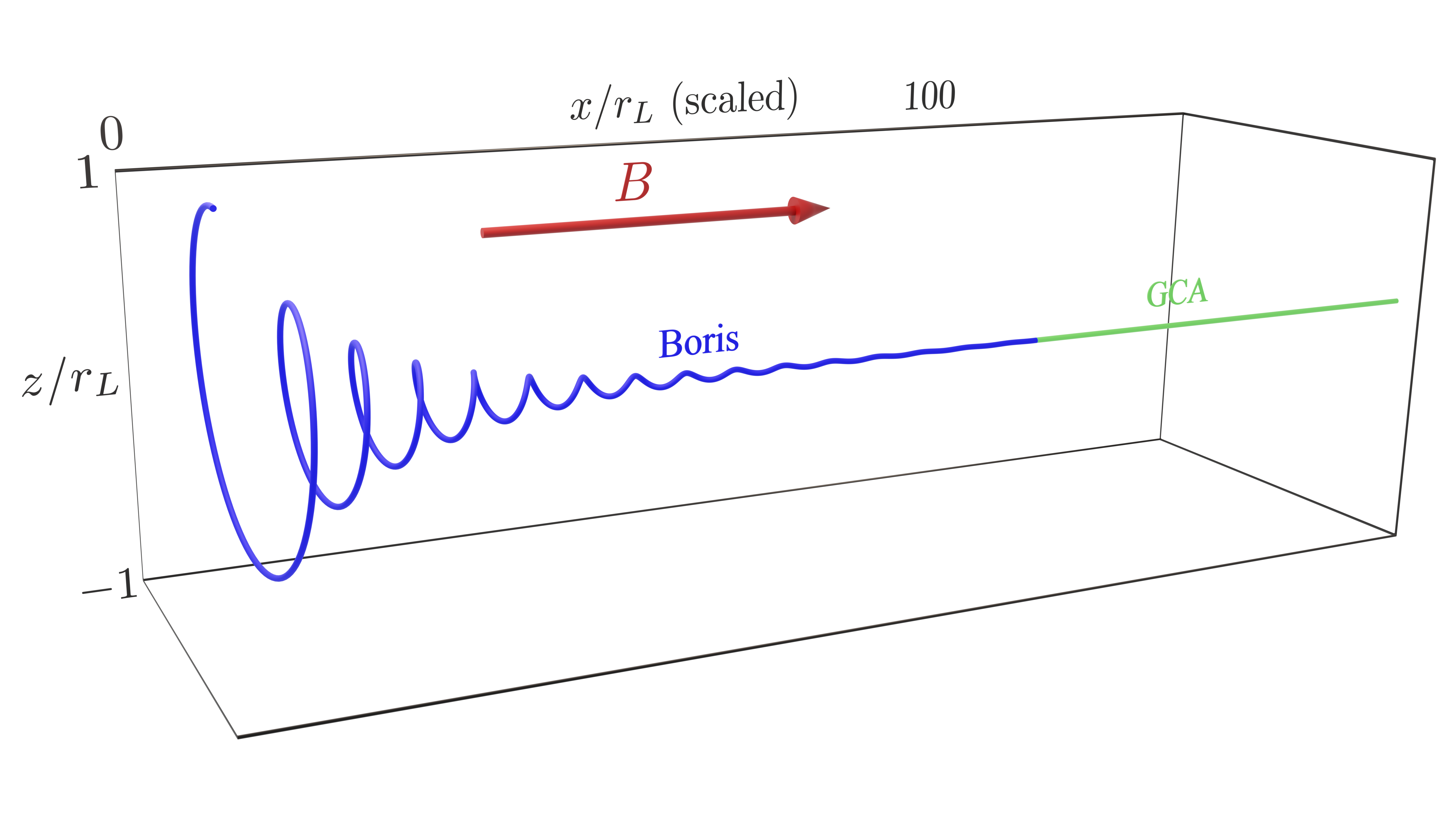}
    \caption{Synchrotron radiation with Boris-GCA coupling. The blue line represents the Boris trajectory, and the green line represents the GCA trajectory.}
    \label{fig:coupled_sync_traj}
\end{figure}

As the particle radiates, its Larmor radius shrinks and eventually reaches a point where it is unresolved by the simulation grid (in this case assumed to have uniform spacing $c\Delta t$, where $\Delta t$ is the integration timestep). The algorithm then switches to GCA and continues integrating the trajectory of the particle gyrocenter. This transition can clearly be seen in figure~\ref{fig:coupled_sync_traj}, where two sections of the trajectory are color-coded in blue and green, integrated with the Boris and the GCA algorithm respectively.

\begin{figure}[htb!]
    \centering
    \includegraphics[width=\columnwidth]{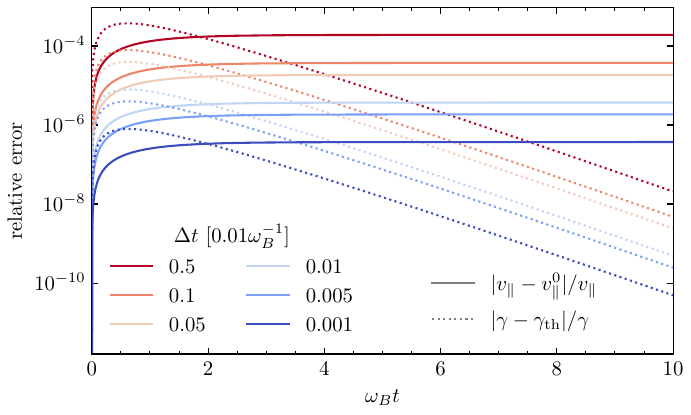}
    \caption{Relative error in $\gamma$ compared to the analytical solution for synchrotron drag.}
    \label{fig:sync_err}
\end{figure}

We then test our implementation of the synchrotron drag force itself, by comparing the evolution of the particle energy and the parallel component of its three-velocity. Multiplying equation~\eqref{eq:syncforce} by the three-velocity of the particle, $\bm{v}$, and setting $\bm{e}=\bm{0}$, we find the energy loss rate: $\dot{\gamma} = - \eta_{\rm rec} \omega_B\gamma^2v_\perp^2 / \gamma_{\rm syn}^2$, where $v_\perp^2 = v^2 - v_\parallel^2$ is the component of the three-velocity perpendicular to the magnetic field. This equation can be solved exactly for an arbitrary pitch angle, and is denoted in the figure with $\gamma_{\rm th}$. On the other hand, multiplying~\eqref{eq:syncforce} by the unit vector in the direction of the magnetic field, we see that the value of $v_\parallel$ does not change during the motion of the particle. 

Figure~\ref{fig:sync_err} shows the relative error in the particle energy and in $v_\parallel$. To ensure that the algorithm converges numerically, we vary the timestep from $5\cdot 10^{-3}\omega_B^{-1}$ to $10^{-5} \omega_B^{-1}$. In all cases, the relative error remains below $0.1\%$, and converges with decreasing timesteps.

\subsection{Motion in curvilinear coordinates}

In this section we present unit tests for verifying the accuracy of the curvilinear pusher described in sections~\ref{GR:Boris} and \ref{GR:GCA} both with and without EM fields. In the first experiment we set flat spacetime, $r_g = 0$ in spherical coordinates $(r,\theta, \phi)$, where $\theta \in [0, \pi)$ and $\phi \in [0, 2\pi)$, and prescribe EM fields. Consequently, we also test the pusher in the GR Kerr metric with no EM fields in spherical coordinates.

\subsubsection{Motion in a rotating monopole}

To test the GR-GCA framework in curvilinear coordinates, we compare particle motion in the Michel monopole~\cite{1973ApJ...180L.133M} solution versus analytically known trajectories.
In this experiment, a proton is initialized at rest at $r=2r_s$, where $r_s$ is the fiducial distance from the polar origin. Electric and magnetic fields are prescribed to follow the Michel rotating monopole solution: $\bm{D}=-\hat{\bm{\theta}}B_s(r_s/R_{LC})(r_s/r)\sin{\theta}$, and $\bm{B} = \hat{\bm{r}}B_s(r_s/r)^2+\hat{\bm{\phi}}D_\theta$, where $B_s$ is the fiducial magnetic field strength, while the parameter $R_{LC}$, which we pick to be $10 r_s$, corresponds to the light cylinder (where $|D_\theta|=|B_r|$ at $\theta=\pi/2$). In these fields, the motion of the particle is described as an $\bm{D}\times \bm{B}$-drift in the radial direction ($\bm{D}\times \bm{B} 
= D_\theta B_\phi \hat{r}$) with the Lorentz factor of the drift increasing with distance (the drift becomes relativistic for $r>R_{LC}$).

\begin{figure}
    \centering  
    \begin{subfigure}[b]{\columnwidth}
        \includegraphics[width=\textwidth]{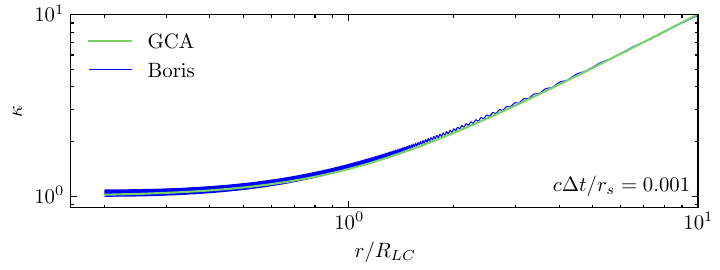}
        \subcaption{Lorentz factor $\kappa$ relating to the $\bm{D} \times \bm{B}$ drift for Boris (blue) and GCA (green) coupled with the GR push }
        \label{fig:michel traj}
    \end{subfigure}
    \begin{subfigure}[b]{\columnwidth}
        \includegraphics[width=\textwidth]{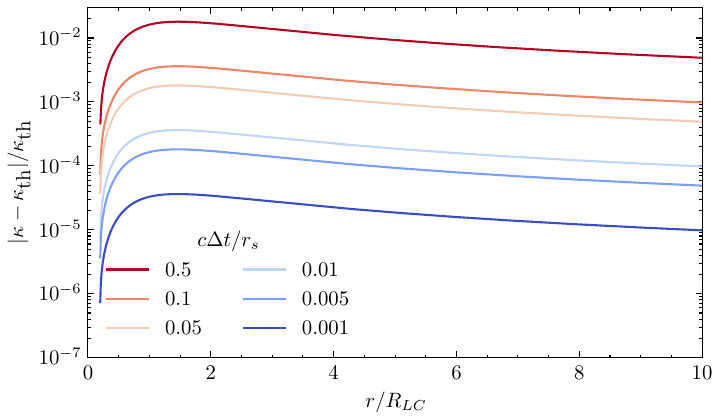}
        \subcaption{Relative error in $\kappa$ compared to analytical solution}
        \label{fig:michel err}
    \end{subfigure}
    \caption{Electron motion in a Michel monopole magnetic field with the Boris pusher and the GCA pusher.}
    \label{fig:michel}
\end{figure}

Figure~\ref{fig:michel} shows the particle $\bm{D} \times \bm{B}$ drift Lorentz factor, $\kappa$, as a function of $r$ from two of our simulations using the conventional Boris algorithm (in blue), and the GCA (in green). We see that both algorithms correctly reproduce the overall trend, with an additional gyration component for the Boris case. Then, in figure~\ref{fig:michel err}, we test the convergence of the GCA algorithm with the timestep, $\Delta t$, ranging in $0.001\mbox{--}0.5~r_s/c$. As expected, the relative error in $\kappa$ converges with the timestep when compared to the analytical solution, $\kappa_{\rm th} = \left(1+(r/R_{LC})^2\right)^{1/2}$.

\subsubsection{Kerr metric}



To test the geodesic step of the pusher in curved spacetime, in the following test we model the motion of an uncharged massive particle around a spinning black hole,  $a=0.995$, with no EM fields. We employ the initial parameters from~\cite{Bacchini:2018zom}, where in Boyer-Lindquist coordinates, the initial four-velocity components of the particle is $u_r=u_\theta=0$, $u_\phi = \mathcal{L}$, where $\mathcal{L}$ is the angular momentum of the particle. Transforming to Kerr-Schild coordinates, which is what we use in our pusher, the four-velocity of the particle becomes: $u_r = (-a\mathcal{L} + 2r_0\mathcal{E})/(a^2 + r_0(r_0-2))$, $u_\theta=0$, and $u_\phi=\mathcal{L}$. The initial position of the particle is $r(0)=r_0$, $\theta(0)=\pi/2$, and $\phi(0)=0$, where the value of $r_0$ is determined by the total energy of the particle, $\mathcal{E}$.

\begin{figure}
    \centering
    \includegraphics[width=\columnwidth]{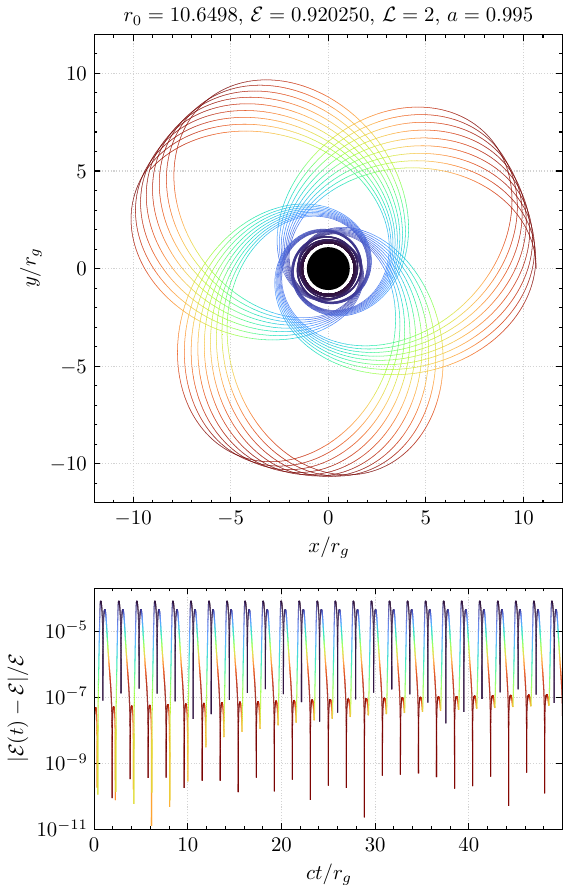}
    \caption{Kerr metric bound-orbit tests for uncharged massive particles. The color varies with radius, with the closest corresponding to purple and farthest to red.}
    \label{fig:three_leaf}
\end{figure}

To produce the three-leaf orbit described in~\cite{Bacchini:2018zom}, we use $\mathcal{E}=0.920250$, $\mathcal{L}=2$, $a=0.995$, which also implies $r_0=10.6498$. Figure~\ref{fig:three_leaf} shows the resulting orbit, where color indicates the distance from $r=0$. The panel below shows the relative error in the total energy $\mathcal{E}(t)$, with color corresponding to the first panel. As expected, the orbit of the particle has three ``leaves'' which precess over time. The relative error is bound below $10^{-4}$ at all times, with closer encounters to the event horizon (blue regions) being the least accurate.




\subsection{Hadronic processes}

In this section we present two experiments testing the drag term due to proton-proton interactions. We limit the discussion to only this process and in Cartesian Minkowski spacetime, since both the $p\gamma$ and \emph{bth} interactions are implemented similarly (with cross sections defined in  appendix~\ref{appendix:cross_sections}), and the procedure can be easily generalized to GR by transforming to the tetrad basis. We study two cases: for static and moving background protons; in the latter case, the incoming proton velocity, the background number density, and the timestep are transformed to the frame of the moving background, where the problem reduces to the former case. 

\subsubsection{$pp$ interaction with a static proton background}

In this test, we model the interaction of a proton moving through a uniform background population of cold protons. The cross section of the interaction is computed as described in section~\ref{sec:hadronic_drag} by fitting the experimental data. We define $\lambda_{pp}^0=(n\sigma_{pp})^{-1}$ by approximating $\sigma_{pp}\approx 50$~mbarn, which gives us the expected energy dependence on time: $\gamma(t) = \gamma_0\exp(-\xi_{pp}ct/\lambda_{pp}^0)$, where $\xi_{pp}\approx 0.17$ is the inelasticity of each collision.

\begin{figure}
\centering
\includegraphics[width=\columnwidth]{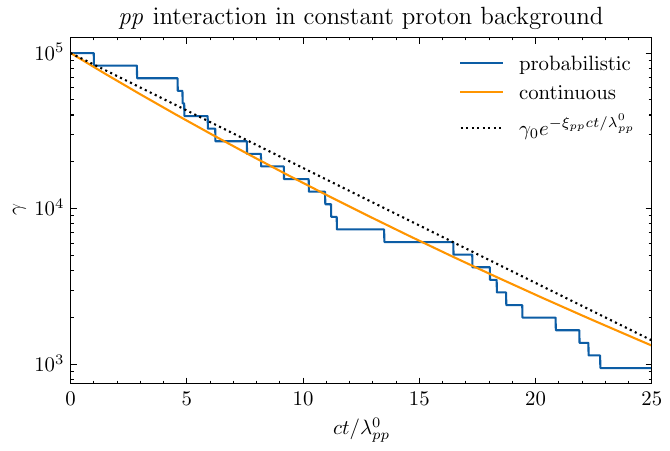}
\caption{Proton gyration around a constant magnetic field through a proton field. The orange trajectory shows the Monte-Carlo simulation scheme, while the blue trajectory depicts the one with continuous loss. The dashed line is the analytical expression for a constant cross section.}
\label{fig:hadronicpp_gyr}
\end{figure}

We model this interaction using both the continuous and the probabilistic implementation of the drag according to equation \eqref{eq:pp_prob_cont}, by starting with $\gamma_0=10^5$, and employing a timestep $c\Delta t / \lambda_{pp}^0=5\cdot 10^{-3}$. Figure~\ref{fig:hadronicpp_gyr} shows the results of this test both for the continuous (orange), and the probabilistic (blue) method, together with the analytic expectation (black dashed). As the timestep, $c\Delta t$, is small-enough compared to the mean free path, both approaches show the same trend, which in turn is close to the expectation from the asymptotic constant cross section approximation. 

\subsubsection{$pp$ interaction with a moving background}

The interaction with a moving background field of protons is tested in a similar way, except that we first transform to the frame of the background. In this experiment, the incoming proton starts with a velocity purely in the $x$-direction, $u_x= 10^3$, while the cold background moves along $y$ with a marginally relativistic bulk velocity $u_y^b=1$ (with a Lorentz factor $\Gamma=\sqrt{2}$). As the proton experiences the drag from the moving background, its total energy drops, until its momentum matches that of the background. Initially, as $u_x\gg \Gamma$, the energy of the particle decreases roughly as in the previous case, $\gamma(t)=\gamma_0 \exp(-\xi_{pp} ct/(\Gamma^2\lambda_{pp}^0))$, where the additional factor of $\Gamma$ accounts for the fact that the effective mean free path is larger in the lab frame compared to the frame of the background flow. 

\begin{figure}
    \centering
    \includegraphics[width=\columnwidth]{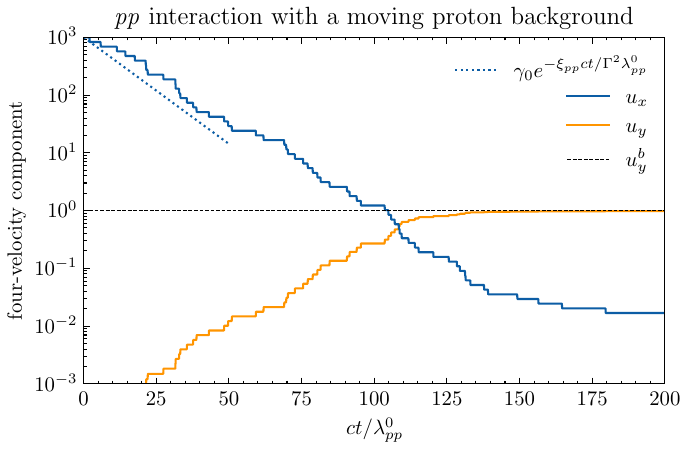}
    \caption{Proton motion through a proton field with a net bulk velocity in the $y$ direction. The blue line represents the $x$-component of the particle 4-velocity, while the orange line represents the $y$ component. The blue dashed line is the analytical expression for a constant cross section (i.e. $\lambda_{pp} = \lambda^0_{pp}$)}
    \label{fig:hadronicpp bulk}
\end{figure}

In figure~\ref{fig:hadronicpp bulk} we show the evolution of the proton four-velocity components perpendicular to the flow (blue), and parallel to it (orange). As expected, the perpendicular component drops exponentially, until the motion becomes marginally relativistic. $u_y$, on the other hand, is initially zero, and rises to the level of $u_y^b$. After about $ct\gtrsim 100\lambda_{pp}^0$, in the frame of the background the proton is static, while in the lab frame it moves with the background flow and experiences almost no collisions.

\section{Astrophysical Applications}
\label{sec:astro}
This paper introduces a comprehensive set of numerical methods that extend and generalize the study of particle acceleration and cooling in magnetized astrophysical environments. While non-thermal emission from astrophysical sources has traditionally been explored using analytical models, supported by insights from kinetic simulations, we offer a more robust set of tools that enhance the investigation of these phenomena within (GR)MHD and potentially (GR)PIC simulations. These tools allow for the tracking of individual high-energy particles and their interactions with the radiation fields and surrounding matter fields. 

These tools allow us to investigate a range of astrophysical processes, including cosmic ray acceleration, the injection and cooling of ultra-high-energy cosmic rays (UHECRs), and the production of neutrinos, $\gamma$-rays, X-rays, and radio emissions across various astrophysical sources. Notably, the longer cooling times expected for hadronic interactions in the environments we study make propagation within (GR)MHD simulations particularly valuable. These cooling times, being longer than those relevant to smaller PIC scales, are more appropriately linked to the largest-scale regions. Some key applications include:

\paragraph{Jets--}
(GR)MHD simulations model the formation and launching of jets in environments such as Active Galactic Nuclei (AGN), $\gamma$-ray bursts (GRBs), and X-ray binaries. This framework enables investigations into particle acceleration \citep{mbarek+19,mbarek+21,wang+23} and neutrino/$\gamma$-ray emission \citep{Mbarek:2022} in these sources. Similar methods have been applied to specific cases, such as Centaurus A, to estimate the neutrino yield at the highest energies and compare predictions with future experimental resolutions \citep{mbarek+24b}. Photodisintegration of heavier elements might be necessary to fully study such sources, but could be easily implemented within our framework (see \cite{Mbarek:2022}).

\paragraph{AGN Disks, magnetospheres, and coronae--}
Near supermassive black holes, extreme environments characterized by intense magnetic and radiation fields, along with various outflows, create ideal conditions for particle acceleration to relativistic speeds, followed by subsequent cooling. The propagation and cooling of pre-accelerated ion populations can be studied through large-scale GRMHD simulations of accretion flows around black holes \citep[e.g.,][]{ripperda+22}, particularly where large current sheets are present \citep{ripperda+20}.

Magnetic flux eruptions, observed in thin radiatively efficient disks \citep[e.g.][]{scepi+22,liska+22}, present promising opportunities for ion acceleration and cooling. In contrast, cosmic ray acceleration in radiatively inefficient accretion disks has been explored in prior studies \citep{Fuj+15,kimura+15,Kim+19}, and our framework provides a robust approach to validate and expand these calculations.

Additionally, black hole coronae offer optimal conditions for ion acceleration, cooling, and subsequent neutrino emission. While localized PIC simulations offer valuable insights into these processes \citep{mbarek+24}, large-scale simulations incorporating both acceleration and cooling effects are essential for a more comprehensive understanding of neutrino emission in these regions \citep[e.g.,][]{fiorillo+24,mbarek+24}.


\paragraph{Binary Neutron Star Mergers--}
The violent conditions during these mergers ---characterized by strong magnetic fields, relativistic jets, and shock waves --- are ideal for accelerating particles to relativistic speeds. Our methods can help to assess, e.g., whether heavy UHECRs can originate from such events \cite{farrar24}, whether neutrinos could serve as multimessenger signals in these scenarios \cite{fang+17,kimura+18,pian21}, or whether gamma-ray bursts can efficiently produce neutrinos \cite{most+24}. 

\section{Conclusion}
\label{sec:conclusions}

This study introduces a comprehensive framework for simulating charged particle propagation, integrating advanced numerical methods with plasma-radiation interactions and particle propagation and acceleration in curved spacetime. A key feature of the framework is a novel Monte Carlo approach that models discrete cooling effects from neutrino-generating hadronic processes, such as proton-proton and photomeson interactions. 

The framework is capable of evolving particles within both static and dynamically evolving electromagnetic and gravitational fields, derived from GRMHD and GRPIC simulations. The method is adaptable to any spacetime, making it applicable to phenomena such as compact object mergers and stellar collapse, and can easily be extended to incorporate other relevant radiation and scattering effects.  

All of the algorithms outlined in this paper, together with the benchmark unit tests we describe, are available publicly on GitHub\footnote{https://github.com/Mynghao/pusher-library}.

\section*{Acknowledgements}

B.R.\ is supported by the Natural Sciences \& Engineering Research Council of Canada (NSERC), the Canadian Space Agency (23JWGO2A01). B.R. and L.S. are supported by a grant from the Simons Foundation (MP-SCMPS-00001470). B.R. acknowledges a guest researcher position at the Flatiron Institute, supported by the Simons Foundation.
F.B.\ acknowledges support from the FED-tWIN programme (profile Prf-2020-004, project ``ENERGY'') issued by BELSPO, and from the FWO Junior Research Project G020224N granted by the Research Foundation -- Flanders (FWO). 
The computational resources and services used in this work were partially provided by facilities supported by the VSC (Flemish Supercomputer Center), funded by the Research Foundation Flanders (FWO) and the Flemish Government – department EWI and by Compute Ontario and the Digital Research Alliance of Canada (alliancecan.ca).

\appendix

\section{Boris and GCA pushers in Minkowski spacetime}

\label{appendix:boris_gca}

In a leap-frog scheme, the timestep starts with the velocities $\bm{u}^{n-1/2}$ at time $n-1/2$, and coordinates $x^{n}$ at time $n$, together with the time-aligned electromagnetic fields, $E^n$ and $B^n$. The procedure for the Boris algorithm (see, e.g., \cite{Birdsall&Langdon,ripperda2018a}) can be written in the following form



\begin{equation}
\begin{aligned}
    \bm{u}^- &= \bm{u}^{n-1/2} + k\bm{D}^{n}, \\
    \bm{u}^+ &= \bm{u}^- + \frac{2}{1+w^2} (\bm{u}^- + \bm{u}^- \times \bm{w}) \times \bm{w}, \\
    \bm{u}^{n+1/2} &= \bm{u}^+ + k\bm{D}^{n}, \\
    \bm{x}^{n+1} &= \bm{x}^n + \frac{\bm{u}^{n+1/2}\Delta t}{\sqrt{1+|\bm{u}^{n+1/2}|^2/c^2}},
\end{aligned}
\end{equation}
where $k = {q}\Delta t/(2mc)$, and $\bm{w} = k\bm{B}^{n}/\sqrt{1+(\bm{u}^-)^2/c^2}$. 

When the fiducial Larmor radius of the particle, defined as $u_\perp mc/(eB)$, is smaller than a certain threshold value (assuming, e.g., $D<B$), we can reduce the particle motion to that of its guiding center. The corresponding discretized equations on the position of the guiding center, $\bm{x} = \bm{R}$, and the velocity of the particle parallel to the local magnetic field, $u_\parallel$, will take the following form:

\begin{equation*}
    \begin{aligned}
    \frac{u_\parallel^{n+1/2} - u_\parallel^{n-1/2}}{\Delta t} &= \frac{q}{mc} D_\parallel^n, \\
    \frac{\bm{R}^{n+1}-\bm{R}^n}{c\Delta t} &= 
    \bar{\bm{v}}_\parallel^{n+1/2} + \frac{\bm{v}_D^n + \bm{v}_D^{n+1}}{2},
    \end{aligned}
\end{equation*}
where
    
\begin{equation*}
    \begin{aligned}
    \bar{\bm{v}}_\parallel^{n+1/2}&\equiv \frac{u^{n+1/2}_\parallel}{2}\left(\frac{\bm{\hat{B}}^{n}}{\gamma(\bm{R}^{n},u_\parallel^{n+1/2})} + \frac{\bm{\hat{B}}^{n+1}}{\gamma(\bm{R}^{n+1},u_\parallel^{n+1/2})}\right), \\
    \gamma(\bm{R},u_\parallel) &\equiv \kappa \left(1+ u_\parallel^2 + 2\mu |\bm{B}(\bm{R})|\kappa(\bm{R})/(mc)^2\right)^{1/2}.
    \end{aligned}
\end{equation*}

\noindent Here, $\hat{\bm{B}}$ is the unit vector in the direction of the $\bm{B}$-field, $D_\parallel \equiv \bm{D}\cdot\hat{\bm{B}}$, $\bm{v}_D \equiv \bm{w}_D(1-\sqrt{1-4|\bm{w}_D|^2})/(2|\bm{w}_D|^2)$, $\bm{w}_D\equiv \bm{D}\times \bm{B}/(|\bm{D}|^2 + |\bm{B}|^2)$ and $\kappa \equiv 1/\sqrt{1-|\bm{v}_D|^2/c^2}$. $u_\parallel^{n+1/2}$ can be found explicitly, using the electric field at position $\bm{R}^n$. On the other hand, to find $\bm{R}^{n+1}$ one needs to solve the equation implicitly. In our case, we use a fixed point iteration technique, with the relative error on $k$-th iteration defined simply as $\delta_k = |(\bm{R}^{k} - \bm{R}^{k - 1}) / \bm{R}^k|$. The details to derivation and implementation of the GCA are further explored in \cite{Ripperda:2017, Bacchini:2020, Zou:2023}. 


\section{Photon field rest frame}
\label{appendix:photon_field_trans}

When considering hadronic interactions with an anisotropic background photon field, it is more convenient to transform to the frame where photons are isotropic and perform all the relevant momentum updates in that frame before transforming back. Below we demonstrate how this special frame can be found for a specific case where the photons have a net flux in the direction $x$; this approach can be trivially extended to other cases. 

The stress-energy tensor of the anisotropic radiation field can be written as $T^{00}=U$, $T^{0i}=T^{i0}=\hat{x}^i f U$, and $T^{ii}=\mathrm{diag}(P_\parallel, P_\perp, P_\perp)$, where $U$ is the energy density of the radiation, $P_\parallel$ and $P_\perp$ are the scalar pressure components along the anisotropy (in $x$) and perpendicular to it (in $y$ and $z$), while the dimensionless parameter $f$ encodes the level of anisotropy: $|f|\leq 1$, with $f=0$ corresponding to a perfectly isotropic distribution. One can show that, by transforming to a reference frame defined by a three-velocity (in the $x$ direction) $v_\epsilon = (1-f^*)/f$, where $f^*\equiv \sqrt{4-3f^2}-1$, the stress-energy tensor will become diagonal: $T^{\mu'\nu'}=\mathrm{diag}(U',U'/3,U'/3,U'/3)$, with $U'=Uf^*$ being the energy density of photons in the new frame.

\section{Cross sections for hadronic interactions} \label{appendix:cross_sections}

\begin{figure}
    \centering
    \includegraphics[width = \columnwidth]{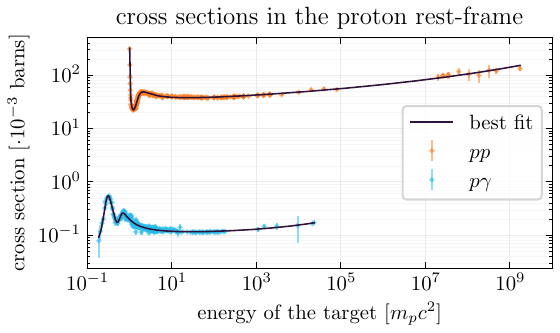}
    \caption{Cross section fit as a function of target proton and photon energies for $pp$ and $p\gamma$ interactions, respectively. Orange data points indicate $pp$ processes, and blue points indicate $p\gamma$.}
    \label{fig:cross_section_fits}
\end{figure}

We use experimental data to fit energy-dependent cross sections for the $pp$ and $p\gamma$ interactions, extracted from the Particle Data Group \cite{ParticleDataGroup:2020ssz}. The data points we used for both of these cases are shown in figure~\ref{fig:cross_section_fits} with orange and blue errorbars respectively. The horizontal axis of the plot corresponds to the energy of the target particle (a background proton or a photon) in the rest-frame of the incoming proton. Black lines show our best-fits using rational functions described below. For the Bethe-Heitler, we directly fit the combination of $\xi_{bth}\sigma_{bth}$, given analytically by equation~3.4 in \cite{1992ApJ...400..181C}. Fitting a polynomial allows for faster computation of the otherwise complex expression for the cross section. Polynomial expressions we used are as follows,

\begin{gather}
   \log_{10}\sigma_{pp} = s_{pp} \frac{\sum_{m} \alpha_m (\log_{10}\bar{\gamma}_p)^m}{\sum_{n} \beta_n (\log_{10}\bar{\gamma}_p)^n}, \\
   \log_{10}\sigma_{p\gamma} = s_{p\gamma} \frac{\sum_{m} \alpha_m (\log_{10}\bar{\epsilon})^m}{\sum_{n} \beta_n (\log_{10}\bar{\epsilon})^n},\\
   \xi_{bth}\sigma_{bth} = s_{bth} \sum_{m} \alpha_m (\log_{10}\bar{\epsilon})^m,
\end{gather}
where $s_{pp,p\gamma,bth}$ are normalization factors, while $\alpha_m$ and $\beta_n$ are the polynomial coefficients. Numerical values for all these coefficients are given in the table~\ref{table:coefficients}. Here, $\bar{\epsilon}$ and $\bar{\gamma}_p$ are, respectively, the characteristic energy of the background photons and protons in the rest-frame of the incoming proton.

\begin{figure}[htb!]
    \centering
    \includegraphics[width = \columnwidth]{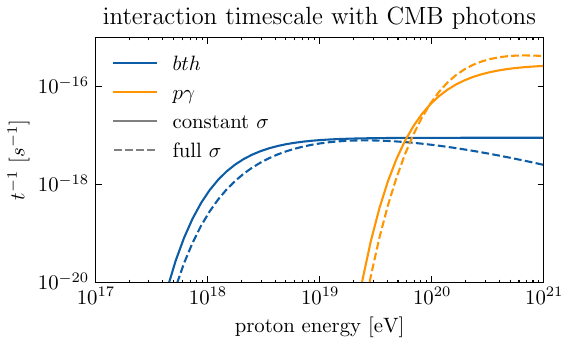}
    \caption{$p\gamma$ and bth cooling timescales, for constant cross sections (solid) and full cross sections (dashed).}
    \label{fig:bth_pg_timescales}
\end{figure}

Figure~\ref{fig:bth_pg_timescales} shows the cooling timescales for both the Bethe-Heitler (blue) and photomeson (orange) interactions, described by equations~\eqref{eq:pg_cool}, and \eqref{eq:bth_cool}, with cosmic microwave background (CMB) photons as a function of the proton energy. We compare our polynomial fits for both processes (dashed lines) with an approximate analytic expression derived assuming a constant cross section given by equation~\eqref{eq:pg_cool_time_CMB}. 


\begin{table}[h]
\centering
\begin{tabular}{|c||c|c||c|c||c|}
\hline
 & \multicolumn{2}{c||}{$pp$} & \multicolumn{2}{c||}{$p\gamma$} &
 \multicolumn{1}{c|}{$bth$} \\
\hline
\hline
 & \multicolumn{2}{c||}{$s_{pp}$} & \multicolumn{2}{c||}{$s_{p\gamma}$} &
 \multicolumn{1}{c|}{$s_{bth}$} \\
\hline
& \multicolumn{2}{c||}{$1.693$} & \multicolumn{2}{c||}{$-1.155$} &
\multicolumn{1}{c|}{$10^{5}$} \\
\hline
$m$, $n$ & $\alpha_{pp, m}$ & $\beta_{pp, n}$ & $\alpha_{p\gamma, m}$ & $\beta_{p\gamma, n}$ &
$\alpha_{bth, m}$\\
\hline
$0$ & $0.004166$ & $0.002128$ & $0.2217$ & $0.3646$ & $15.48$\\
$1$ & $0.1783$ & $0.2875$ & $2.459$ & $3.634$ & $-20.23$\\
$2$ & $-1.404$ & $-2.043$ & $10.45$ & $13.78$ & $11.42$\\
$3$ & $6.602$ & $7.135$ & $17.81$ & $21.06$ & $-4.843$\\
$4$ & $-0.1341$ & $-0.3566$ & $8.251$ & $-309.1$ & $-0.3493$\\
$5$ & - & - & $-4.889$ & $-5.062$ & $3.600$\\
$6$ & - & - & $0.5046$ & $0.469$ & $-0.2164$\\
$7$ & - & - & - & - & $-1.556$\\
$8$ & - & - & - & - & $-1.601$\\
$9$ & - & - & - & - & $0.3862$\\
$10$ & - & - & - & - & $0.06724$\\
$11$ & - & - & - & - & $-0.03500$\\
$12$ & - & - & - & - & $-0.007843$\\
\hline
\end{tabular}
\caption{Coefficients $\alpha_i$ and $\beta_i$ for $pp$, $p\gamma$, and $bth$ cross section fits, kept to 4 significant figures.}
\label{table:coefficients}
\end{table}

\bibliography{references}

\end{document}